\documentclass[16pt]{article}
\usepackage[table,xcdraw]{xcolor}
\usepackage{amsmath}
\usepackage{graphicx}
\usepackage{caption}
\usepackage[export]{adjustbox}
\usepackage{subcaption}
\usepackage[normalem]{ulem}
\useunder{\uline}{\ul}{}
\usepackage{lscape}
\usepackage{csquotes}
\usepackage{bm}
\usepackage{rotating}
\usepackage{apacite}
\usepackage{multicol} 
\usepackage{appendix}
\usepackage{natbib}
\usepackage{hyperref}
\usepackage{doi}
\newcommand{\vdot}{\cdot}

\newcommand{\pdif}[2]{ \frac{\partial #1}{\partial #2}}
\newcommand{\na}{ \nabla}  
\usepackage{amssymb}

\title{A continuum model for the elongation and orientation of \\{Von Willebrand Factor} with applications in arterial flow}
\author{E.F. Yeo$^{1*}$, J.M. Oliver$^2$, N. Korin$^3$, S.L. Waters$^2$\\
\small$^1$ Department of Applied Mathematics, University College London\\
\small$^2$ Mathematical Institute, University of Oxford\\
\small$^3$ Biomedical Engineering, Technion\\
\small $^*$ Corresponding author contact: edwina.yeo.14@ucl.ac.uk}

\begin{document}

\maketitle
\begin{abstract}
The blood protein Von Willebrand Factor (VWF) is critical in facilitating arterial thrombosis. At pathologically high shear rates the protein unfolds and binds to the arterial wall, enabling the rapid deposition of platelets from the blood. We present a novel continuum
model for VWF dynamics in flow based on a modified viscoelastic fluid model that incorporates
a single constitutive relation to describe the propensity of VWF to unfold as a function of the
scalar shear rate. Using experimental data of VWF unfolding in pure shear flow, we fix the
parameters for VWF’s unfolding propensity and the maximum VWF length, so that the protein is half unfolded at a shear rate of approximately $5,000\,\text{s}^{-1}$. We then use 
the theoretical model to predict VWF’s behaviour in two complex flows
where experimental data is challenging to obtain: pure elongational flow and stenotic
arterial flow.

 In pure elongational flow, our model predicts that VWF is 50\% unfolded at
approximately $2,000\,\text{s}^{-1}$, matching the established hypothesis that VWF unfolds at lower shear
rates in elongational flow than in shear flow.  We demonstrate the sensitivity of this elongational flow prediction to the value of maximum VWF length used in the model, which varies significantly across experimental studies, predicting that VWF can unfold between $600\text{ - }3,200\,\text{s}^{-1}$ depending on the selected value. Finally, we examine VWF dynamics in a range of idealised arterial stenoses, predicting the relative extension of VWF in elongational flow structures in the centre of the artery compared to high-shear regions near the arterial walls.

 \end{abstract}
\section{Introduction}

Coronary heart disease is characterised by the formation of plaque on the walls of arteries leading to the muscles of the heart, restricting blood flow \citep{casa2017thrombus}. Plaque rupture can occur due to injury or vessel collapse and a blood clot then rapidly forms to repair the damaged wall \citep{arroyo1999mechanisms}. Blood clot formation in arterial conditions, known as high-shear thrombosis, is facilitated by the shear-sensitive blood protein Von Willebrand Factor (VWF). This protein has platelet binding sites along its length and is tightly coiled at normal levels of the fluid shear rate. At pathologically high shear rates, which occur as the blood flow accelerates over the plaque deposit, the protein unfolds and facilitates the formation of a platelet-based clot in the artery.  

Von Willebrand Factor is a large protein which naturally exists in the blood as a chain of repeating units known as dimers. VWF can be composed of between two and eighty dimers, and the proteins with the most dimers play the most dominant role in haemodynamics \citep{ Furlan1996,Sadler1998}.  VWF shape and length in flow was first revealed in 1996 when atomic-force microscopy was used to demonstrate that the protein only unfolds at shear stresses greater than 35\,dyne/cm$^2$, which is equivalent to a shear rate of $3,500\,\text{s}^{-1}$ assuming the suspending fluid has the viscosity of water \citep{siediecki1996shear}.
VWF unfolding has since been characterised experimentally for a range of shear rates  \citep{Lippok2016,bergal2022conformation,Schneider2007}. All of these works find that VWF unfolds in shear flow with shear rates exceeding approximately $5,000\,\text{s}^{-1}$. However, there is significant variation in the maximum length of the protein at high shear rates obtained in the two studies which directly measure VWF extension: \cite{Schneider2007} found that VWF obtains a maximum extension of 15$\,\mu$m in contrast to the value of 0.17$\,\mu$m found by \citep{bergal2022conformation}.
This variation was attributed in \cite{bergal2022conformation} to the blurring of VWF in the images obtained at high flow speeds in \cite{Schneider2007}. This disparity in the experimental measurement of VWF length in flow adds uncertainty to the parameters used in theoretical models and therefore to their predictions.

The flow within diseased arteries has complex combinations of regions of shear flow, as well as flow constriction and expansion around narrow regions known as stenoses. However, the small size of VWF combined with the extremely high flow speeds required to unfold the protein present significant experimental challenges. The limited experimental studies of VWF have all studied its behaviour in simpler setups for instance using pure shear flow, which does not reflect the complexities of physiological arterial flow (see Section \ref{section-flow}). 
\cite{Fu2017} avoided the problem of tracking the protein at high speed by tethering VWF to a wall. This study yielded extensive quantitative data on the VWF mechanics, however, it also demonstrated that VWF behaviour differs when the protein is tethered to a wall compared to when it is free to move. Specifically, \cite{Fu2017} found that VWF unfolds at low shear rates when tethered, which is likely due to the fact that the protein cannot resist extension by rotating when it is tethered to a wall. As a result of these experimental challenges, the behaviour of VWF 
in the complex flows that occur within arteries is not well understood. Predicting VWF's dynamics within arteries is critical to understanding, and ultimately treating, thrombosis in clinically relevant scenarios. 

In this paper, we present a novel theoretical model which describes VWF dynamics, we explore VWF dynamics using this model in steady simple two-dimensional flows and in steady stenotic arterial flow for Reynolds numbers up to 500, where the Reynolds number is based on the maximum vessel radius. This captures the range of flow rates seen in small arteries near the heart \citep{mahalingam2016numerical}. In this range, steady flow will remain laminar, since the threshold for turbulence is $Re=1000$ for a stenosed pipe which is 50\% obstructed (referred to as a 50\% stenosis) \citep{ahmed1983flow,mahalingam2016numerical}. 
We now summarise the fluid flow 
within stenosed arteries which we will see underpins VWF behaviour in flow. 

\subsection{Flow structure within diseased arteries}\label{section-flow}

The action of a fluid flow $\bm{u}$ on suspended proteins can be described locally by the deformation gradient $\na \bm{u}$. The deformation gradient can be split into symmetric and antisymmetric components:
\begin{align}
    \na\bm{u}=\frac{1}{2}\left(\na\bm{u}+\na\bm{u}^T\right)+\frac{1}{2}\left(\na\bm{u}-\na\bm{u}^T\right)=\bm{D}+\bm{W},
\end{align}
where $\bm{D}$ and $\bm{W}$ are the rate of strain tensor and the rotation tensor and describe local extension and local rotation, respectively. The magnitude of strain 
 and rotation can be quantified through the scalar shear rate and scalar rotation rate defined as follows 
 \begin{align}
   \dot{\gamma}=\sqrt{2\bm{D}:\bm{D}}, \quad    \dot{\omega}=\sqrt{2\bm{W}:\bm{W}},\end{align}
   where $:$ denotes the double dot product. 

\begin{figure}[t!]
    \centering
    \includegraphics[width=\textwidth]{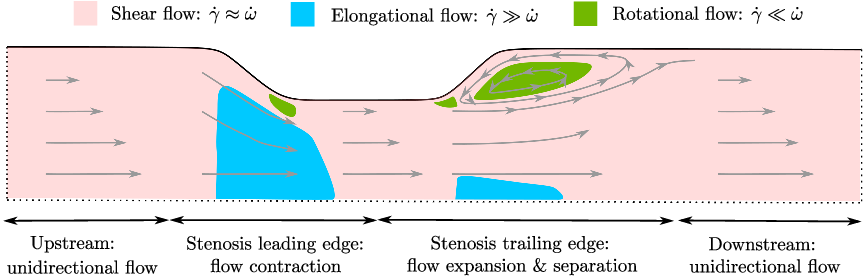}
    \caption{Sketch of the flow structures within a two-dimensional slice of a 3D axisymmetric stenosed artery based on our numerical simulations at $Re=500$ with a 50\% stenosis. Dotted lines illustrate the inlet (left), outlet (right) and the centre of the pipe (bottom). Four regions of key flow behaviour are labelled below.    Illustrative streamlines are shown by grey arrows. Three key flow structures are highlighted by colours at their respective locations within the artery. 
    Shear flow (pink), which is an approximately equal combination of rotational and elongational flows, occurs near the vessel walls and away from the stenosis where flow is unidirectional. 
    Elongational flow (blue) occurs away from the wall at the leading edge of the stenosis, and to a lesser extent at the trailing edge. 
    Rotational flow (green) occurs predominantly in the recirculation zone behind the stenosis, although small regions of rotation occur close to the vessel wall at the leading and trailing edge of the stenosis. }
    \label{fig:flow_structure}
\end{figure}

Stenotic arterial flow is a complex combination of elongational and rotational flows \citep{casa2017thrombus, rana2019shear}. Upstream of the stenosis, the flow is predominantly in the axial direction (assuming that the vessel is not significantly curved) and is categorised as shear flow in which $\dot{\gamma}\approx \dot{\omega}$. The flow accelerates as it reaches a constriction in the vessel leading to an increase in the fluid velocity and the shear rate. We refer to this region as the leading edge of the stenosis. As the flow contracts, for sufficiently steep stenoses, the radial component of the flow becomes comparable to the axial flow, this leads to a region of elongational flow which is defined by $\dot{\gamma}\gg\dot{\omega}$. Just downstream of the stenosis, a region we refer to as the trailing edge of the stenosis, a closed recirculation zone can form in which axial and radial flows are significant. In this region, the flow is rotational, which is defined by $\dot{\gamma}\ll \dot{\omega}$.
On both the trailing and leading edge of the stenosis close to the wall, there are small regions of rotational flow as the flow bends significantly to accommodate the stenosis geometry. Far downstream the flow relaxes back to unidirectional. Fig.\,\ref{fig:flow_structure} illustrates the locations of these flow structures, based on our numerical simulations at $Re=500$ with a 50\% stenosis. Details of the simulation method and fluid flow boundary conditions are listed in Section \ref{section-artery}.

To study these complex flow effects in isolation, theoretical and experimental studies often use idealised simple flows for instance; pure elongational flow, in which $\dot{\omega}=0$; pure rotational flow, in which $\dot{\gamma}=0$; and pure shear flow which has exactly equal parts elongation and rotation so that $\dot{\omega}=\dot{\gamma}$. We use these idealised flows to examine the behaviour of our theoretical model in Section \ref{results-behaviour}, and we use pure shear flow to parameterise our model compared to experimental studies of VWF in Section \ref{param-section}. We now summarise the known behaviour of VWF in different flow structures.

\subsection{VWF behaviour in flow and existing theoretical models}\label{section-vwf-behaviour}

The elongational flow within stenotic arteries has been proposed as a key mechanism in VWF's ability to rapidly unfold \citep{casa2017thrombus,Sing2010}.
 In experimental studies using pure elongational flow, proteins and polymers similar to VWF fully unfold at lower values of shear rate than in pure shear flow \citep{babcock2003visualization,smith1999single}; theoretical models predict this behaviour also occurs for VWF in suspension \citep{Sing2010}. However, all experimental studies of VWF unfolding both in free flow and tethered use shear flow, since tracking and imaging proteins in suspension at high shear rates is less challenging in unidirectional flows \citep{bergal2022conformation,Fu2017,Schneider2007}. No experimental studies to date have examined VWF dynamics in pure elongational flow. As a result, the hypothesis that there is also a lower shear rate threshold of unfolding in pure elongational flow for VWF has not yet been tested \textit{in vitro}.
Furthermore, in elongational flow regions where $\dot{\gamma}\gg\dot{\omega}$, it is unclear how much the shear rate must exceed the rotation rate for the proposed rapid unfolding to occur.
\cite{babcock2003visualization} examined DNA molecules in elongational flow and determined that DNA unfolds more easily if the difference between the shear rate and rotation rate divided by the total rotation and shear, $(\dot{\gamma}-\dot{\omega})/(\dot{\gamma}+\dot{\omega})$, exceeds 0.0048. However, this threshold has not yet been characterised for VWF.

Mathematical models can examine VWF's dynamics in flows that are challenging to generate \textit{in vitro}, namely elongational flows. Existing mathematical models of VWF are predominantly discrete models which describe the protein as a chain of beads and springs. The spring coefficients 
can then be parameterised so that the model predicts VWF unfolding at approximately 5,000$\,\text{s}^{-1}$ in pure shear flow to match experimental data. However these bead and spring models of VWF predict a wide variety of unfolding thresholds in pure elongational flow: 500$\,\text{s}^{-1}$ \citep{Sing2010}, 2,400$\,\text{s}^{-1}$ \citep{nguyen2021unraveling}, 2,500$\,\text{s}^{-1}$ \citep{kania2021predicting} and 3,500$\,\text{s}^{-1}$ \citep{dong2019mechano}.  Discrete models can also examine VWF's interactions with red blood cells or platelets in flow as part of the thrombosis cascade. For instance,  \citep{Rack2017} demonstrate that the proteins remain globular in the centre of the vessel which enables the protein to travel to the edge of the vessel more easily since collisions with red blood cells displace globular proteins further than the unfolded proteins. \cite{liu2022sipa} examine the formation of small platelet-VWF aggregates, and predict the required protein concentration and length to initiate aggregation.

Discrete models of VWF can be characterised with \textit{in vitro} data and offer insights into protein mechanics. However, these models can only accommodate a limited number of proteins and their interactions during thrombosis before their numerical solution becomes demanding. 
An alternative approach is to employ continuum models that examine the dynamics of a large number of constituents such as platelets, red blood cells and proteins, together with their role in thrombosis \citep{Wu2020,Du2020,leiderman2011grow}.  To explicitly incorporate VWF into these models, a continuum description for VWF able to describe the protein's dynamics in complex, evolving arterial flow is required.

VWF dynamics are modelled using a continuum framework in \cite{zhussupbekov2021continuum}. The authors use a two-species model where VWF exists in one of two binary states: either fully unfolded or completely globular. Each species is tracked using an advection-diffusion equation. The unfolding rate which moves proteins from the globular category to the unfolded category is modelled by first classifying the local flow as approximately pure shear, pure elongational or pure rotational, then prescribing unfolding rates in each case. The unfolding of VWF in shear flow uses the empirically determined unfolding rate of \cite{Lippok2016}. In regions defined as elongational flow, according to the DNA threshold of \cite{babcock2003visualization}, the authors use a constant unfolding rate. No unfolding occurs in rotational flow. This model predicts that, in a stenotic flow, a significant number of VWF molecules are unfolded both close to the stenosis wall in the shear flow region and away from the wall due to elongational flow regions. This model was then incorporated into a thrombosis model where shear-flow-induced VWF unfolding near the wall was shown to match the location of thrombus formation \textit{in vitro} \citep{ZHUSSUPBEKOV20224033}.  This is the first work to include an explicit description of unfolding VWF in a continuum model. Other studies include VWF by increasing the phenomenological binding rate between platelets and the vessel wall as a function of shear rate or elongation rate \citep{Du2020,Sorensen1999, Wu2020}.

In this paper, we present a novel continuum model for VWF that predicts the length and orientation of the protein in varying flow structures. Our model describes VWF length and orientation continuously, allowing examination of cases where VWF only partially extends which is vital to examine thrombosis at shear rates marginally outside of the normal range.
Our model does not split the local flow into discrete categories. Instead, our model encodes the flow structure through the deformation tensor and can therefore describe VWF dynamics in shear, elongational and rotational flows and combinations of these in three dimensions using a single unfolding propensity.
This unfolding propensity function can be parameterised using experimental data from pure shear flow which eliminates the need to use data from other proteins as in \citep{zhussupbekov2021continuum} which may be inaccurate for VWF.  Crucially, this allows us to predict the protein unfolding throughout the full range of flow types that occur within diseased arteries. The accuracy of these predictions relies on the corresponding accuracy of our model parameters.
In this paper, we quantify how model predictions change depending on the selected value of the parameter for which the experimental measurements are the most uncertain: the maximum length VWF can reach in flow.

It is important to note that VWF has been shown to demonstrate hysteresis, whereby the time taken for the protein to relax back to its natural length following the removal of flow is much longer than the time taken to unfold when the flow is applied. \cite{Fu2017} demonstrated that tethered molecules unfold over approximately $0.01$\,s when the flow is turned on and require approximately $1$\,s to return to their natural length. The time required to travel the length of the coronary artery can be estimated to be approximately 1\,s (based on an arterial length of 10 cm, and a velocity of $0.1$ m/s \citep{Grief2005}). However, the time required to pass a typical stenosis is approximately 0.1\,s (based on a 1.7 cm stenosis and a 0.16 m/s pathological velocity \citep{elhfnawy2019stenosis,zafar2014measurement}. This means that the proteins could remain partially unfolded in the region downstream of the stenosis. In this paper, we do not consider VWF hysteresis, but it is a valuable extension discussed in Section \ref{section-discussion}.

The paper is structured as follows. First, in Section \ref{section-VWF-model}, we present the mathematical model which is derived from an existing viscoelastic fluid model. In Section \ref{section-artery} we present an idealised arterial stenosis flow setup. In Section \ref{results-behaviour}, we explore the predictions for VWF behaviour in pure shear flow and pure elongational flow, we do not examine pure rotational flow as VWF does not extend in this regime. In Section \ref{param-sensitive} we determine the sensitivity of this elongational flow prediction to the value of maximum VWF length used in the model, which varies significantly between the experiments of \citep{Schneider2007} and \citep{bergal2022conformation}.
In Section \ref{results-artery}, we explore the mechanistic insight that our model can provide in the complex flow regimes inside arteries through direct numerical simulations in a range of idealised stenoses. The flow consists of predominantly shear flow near the stenosis wall and regions of predominantly elongational flow at the leading edge of the stenosis. We select the maximum VWF extension as found in \citep{Schneider2007} and show that the model can predict the relative extension of VWF in the elongational flow structures in the centre of the artery compared to high shear regions near the arterial walls. For this value of maximum VWF extension, we find that VWF is most extended, and therefore most reactive with platelets, in the shear flow close to the stenosis wall. We conclude in Section \ref{section-discussion} by discussing the implications of these predictions, how they can be used to examine VWF's role in arterial thrombosis and highlighting the limitations of the model.

\section{Methods}\label{section-methods}

\subsection{VWF model }\label{section-VWF-model}
We model blood, which contains VWF, using the Navier-Stokes equations and a modified Finitely Extensible Nonlinear Elastic model with the Peterlin spring closure (FENE-P) in the limit where the contribution to the fluid stress from the suspended VWF molecules is negligible. The relative scale of the protein stress compared to the stress of the suspending fluid is determined by the ratio $Gd/\mu U$, where $d$ is a reference lengthscale, $U$ is a reference velocity value, $\mu$ is the fluid viscosity and $G=nk_bT$, in which $n$ is the number of proteins per meter squared, $k_b$ is Boltzmann's constant and $T$ is the average temperature. To estimate the number of proteins per meter squared we use the concentration of VWF in the blood (0.055 g/m$^3$) and the protein's molecular weight (between 500 and 20,000 KDa depending on the number of dimers combined) \citep{Furlan1996,peyvandi2011role}. The value of $G$ can then be estimated as between 0.027 and 6.7$\times10^{-4}$\,Pa$^{-1}$. In this paper, we consider arterial flows with fluid velocities between 0.17 and 0.84\,m/s in a 1.5\,mm radius vessel. Hence the maximum value of the ratio $Gd/\mu U$ is approximately $ 0.008$, demonstrating that VWF's contribution to the fluid stress is minimal. As a result, the flow is uncoupled from the VWF dynamics and VWF does not contribute to the overall fluid stress.

We model blood as an incompressible, Newtonian, viscous fluid with velocity $\bm{u}$ and pressure $p$ at time $t$. The flow is
governed by the incompressible Navier-Stokes equations given by\begin{alignat}{2}
    \na\vdot\bm{u}=0,
    \quad
  \rho\left(\pdif{\bm{u}}t+  \bm{u}\vdot\na \bm{u}\right)=-\na p+\mu\na^2\bm{u},\label{stokes}
     \end{alignat}
where the density $\rho$ and viscosity $\mu$ of the blood are taken to be constant. 

We capture the average length and orientation of VWF molecules via the symmetric, rank 2, configuration tensor $\bm{A}.$ 
The components of $\bm{A}$ can be used to describe the protein's extension in each direction which we will demonstrate for simple flows in Section \ref{results-behaviour}.
The trace of $\bm{A}$ is proportional to the average length of the protein squared \citep{Rallison1988} and hence we define the normalised VWF length as
\begin{align}
    \mathcal{L}=\sqrt{\frac{Tr(\bm{A})}{Tr(\bm{I})}},\label{L_def}\end{align} so that when $\mathcal{L}=1$, the protein is at its natural length for which $\bm{A}=\bm{I}$. We define the extension of the proteins to be $\mathcal{E}=\mathcal{L}-1$, which we use to compare model predictions to experiments in Section \ref{param-section}.

The configuration tensor evolves as a FENE-P fluid and is governed by the following equation
\begin{align}\pdif{\bm{A}}t+
\bm{u}\vdot \na\bm{A}-\bm{A}\vdot\na\bm{u}-&(\na \bm{u})^T\vdot\bm{A}=-\frac{1}{\tau(\dot{\gamma})}\left(\bm{f}(\bm{A})\bm{A}-a\bm{I}\right),\label{fene-main}
\end{align}
where ${\tau}$ is the VWF relaxation time, $f(\bm{A})=\text{L}^2/(\text{L}^2-Tr(\bm{A}))$ is the nonlinear spring law which restricts the protein length to be less than a prescribed maximum we denote L, and $a=\text{L}^2/(\text{L}^2-Tr(\bm{I}))$ is a constant which ensures that in the absence of flow $\bm{A}=\bm{I}$ \citep{bird1980polymer}. 
 In \eqref{fene-main}, the left-hand side represents the transport of proteins and the rotational and extensional effects of the fluid flow, while the right-hand side represents the elastic forces which resist extension. 

We model the unfolding of VWF at high shear rates by allowing the VWF relaxation time ${\tau}$ in \eqref{fene-main} to depend on the shear rate $\dot{\gamma}$. This is described through a saturating function of the fluid shear rate as follows
   \begin{align}
    \tau(\dot{\gamma})=\alpha\left(\frac{1}{2}(\tanh(\beta(\dot{\gamma}-\gamma^*))+1)+\delta\right).\label{tau}
\end{align}
The parameter $\gamma^*$ is the shear rate at which VWF relaxation time is half of its maximum value and $\beta$ describes how quickly the relaxation time varies as the shear rate increases.  Large values of $\beta$ correspond to a rapid increase in $\tau$ once the shear rate reaches $\gamma^*$. Finally, the parameters $\alpha$ and $\delta$ fix the minimum and maximum values of the relaxation time to be $\alpha\delta$ and $\alpha(1+\delta)$ respectively.
This nonlinear relaxation time is shown in Fig.\,\ref{vwf_fit_lip}a. 
Examining the left- and right-hand sides of \eqref{fene-main}, VWF extension is driven by fluid gradients, which are proportional to the shear rate $\dot{\gamma}$, and extension is resisted by elastic forces, which are proportional to the inverse relaxation time, $1/\tau$.  The relative size of these two effects is determined by $\dot{\gamma}\tau$, so that if $\dot{\gamma}\tau\ll1$ then elastic forces dominate and the protein remains globular whereas if $\dot{\gamma}\tau\gg1$, fluid extension forces dominate and the protein unfolds. 
In practice, this means that we select the values of the unfolding parameters $\alpha$ and $\delta$, so that $\alpha(1+\delta)\dot{\gamma}\gg1$ for shear rates where we want VWF to unfold and $\alpha\delta\dot{\gamma}\ll1$ at shear rates where VWF remains globular. We detail the parameter selection method in Section \ref{param-section}.

    There are several important points to note when applying this model.  Firstly, the FENE-P model describes dilute suspensions and does not include protein-protein interactions. This means that we cannot model entanglement or protein-protein binding which may be significant in the later stages of thrombosis.  Secondly, the FENE-P equation is derived through mean-field analysis of a collection of microscopic Brownian dumbbells in the absence of walls \citep{bird1980polymer}. However despite this inconsistency, the model is extensively and successfully used for flows in bounded domains.     
    Including boundary effects in viscoelastic models remains an open theoretical challenge, hence in this paper, we use the FENE-P model 
    to describe the dynamics of VWF in bounded flows. We use the model solution at the boundary to describe the length of VWF close to the wall; gaining insight into the protein dynamics where VWF-platelet binding occurs. 
 Finally, we note that our modified FENE-P model cannot predict VWF hysteresis, since \eqref{fene-main} is single-valued for a particular shear rate. Hence the proteins will relax back to their natural length on the same timescale as they unfolded on. We discuss the limitations of these assumptions on model predictions in Section \ref{section-discussion}.

\subsubsection{Parameterisation}\label{param-section} The model parameters required to describe the flow and VWF behaviour according to \eqref{stokes} and \eqref{fene-main} are shown in Table \ref{params}. The VWF unfolding parameters in \eqref{tau}, namely $\alpha$, $\gamma^*$, $\beta$, $\delta$, and the maximum VWF length L are unknown. 

We estimate these parameters, aside from L, by fitting the numerical solution of the FENE-P equation in simple shear flow to the empirical fitting of normalised VWF extension by \citep{Lippok2016}. We use a gradient-based minimiser in \textit{Matlab} to carry out the fitting.
The data of \citep{Lippok2016} provides relative VWF extension only, leaving the maximum VWF length L, unknown.
The value of extension VWF achieves in flow is not well established experimentally; measured VWF maximum extension ranges from twice its natural length to fifteen times its natural length\newline \citep{bergal2022conformation,Schneider2007}. 
To quantify how our predictions for VWF behaviour in pure elongational flow would change as more data becomes available on VWF length, in Section \ref{param-sensitive} we estimate model parameters,  $\alpha$, $\gamma^*$, $\beta$, $\delta$, using the normalised VWF extension by \citep{Lippok2016} for a range of L values from 5 - 100. This corresponds to a maximum extension between 2 and 70 times the protein's natural length. Details of the parameter estimation algorithm are given in Appendix \ref{appendix-param}.

\begin{figure}[t!]
    \centering
        \includegraphics[width=\textwidth]{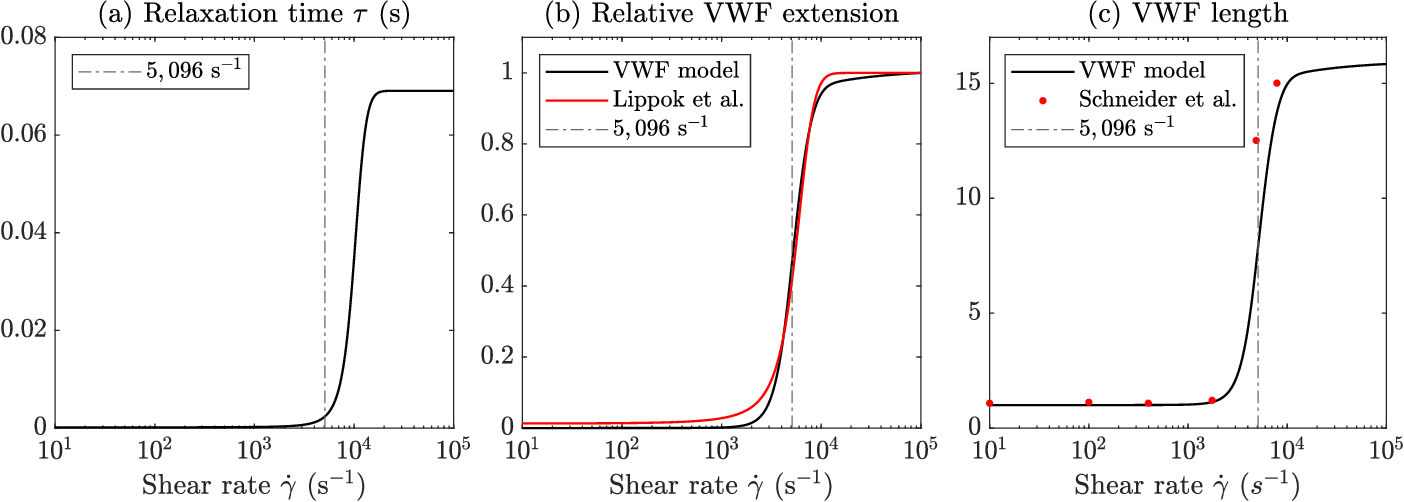}
\caption{(a)  Modified relaxation time of VWF, Eq.\,\eqref{tau}. At low shear rates, the relaxation time is small and the proteins do not extend; at high shear rates, the relaxation time is larger which allows the proteins to extend. (b) Extension of VWF in pure shear flow for varying shear rate compared to the empirical fitting obtained by \citep{Lippok2016} (red); extension is normalised so that the maximum value is one. (c) VWF length $\mathcal{L}$ for varying shear rate in pure shear flow compared to the experimental data of \citep{Schneider2007}.  VWF reaches 50\% of its maximum length at 5,096$\,\text{s}^{-1}$ (dashed-dot line). All subfigures use parameters listed in Table \ref{params} fitted with L$\,=22.6$}\label{vwf_fit_lip}
\end{figure}

For simplicity, in Sections \ref{results-behaviour} and \ref{results-artery} we present VWF dynamics for a fixed value of L, namely L$\,=22.6$, so that the maximum possible extension of VWF matches the value of 15\,$\mu$m obtained by \cite{Schneider2007} when normalised by the globular length of 1\,$\mu$m. The normalised VWF extension in pure shear flow for L$\,=22.6$ is compared to the empirical fitting of \citep{Lippok2016} in Fig.\,\ref{vwf_fit_lip}b. The corresponding VWF length is also compared with the data of \citep{Schneider2007} in Fig.\,\ref{vwf_fit_lip}c. Our fitted model finds that VWF reaches 50\% of its maximum length at 5,096$\,\text{s}^{-1}$ in pure shear flow, this is within 1\% of the unfolding threshold found by \cite{Lippok2016} of 5,122$\,\text{s}^{-1}$.
\begin{table}[t!]
\centering
\begin{tabular}{|l|l|l|l|l|}
\hline
Name                          & Param.       & Value                                                                                                                                       & Units               & Source                                       \\ \hline
Viscosity of blood            & $\mu$        &$    0.0025$                                                                                                                              & Pa s                &     \citep{pries1992blood}              \\ \hline
Density of blood              & $\rho$       & 1050                                                                                                                                              & kg m$^{-3}$         &      \citep{vitello2015blood}                                                            \\ \hline
Extension parameter           & $\alpha$          &                                                                 $    0.069
$                                                                    & s                   &    Fit to \citep{Lippok2016}     \\ \hline
Extension parameter           & $\beta$          &                                                                 $3.44\times10^{-4}$                                                                    & s                   &    Fit to \citep{Lippok2016}     \\ \hline
Extension parameter           & $\delta$          &                                                                                     $9.70\times10^{-4}$                                            & -                   &                                          Fit to \citep{Lippok2016}       \\ \hline
Extension parameter           & $\gamma^*$   &                                                                                          $1.0\times10^{4}$                                                & $\text{s}^{-1}$                   &                                       Fit to \citep{Lippok2016}        \\ \hline
Maximum VWF length   &  L        &                                      \begin{tabular}[c]{@{}l@{}}S.\ref{results-artery} \& \ref{results-behaviour}: 22.6  \\ S.\ref{param-sensitive} : $5-100$                                                     \end{tabular}                                     & -                   &           \begin{tabular}[c]{@{}l@{}} \citep{Schneider2007}  \\ -\end{tabular}                                                          \\ \hline
\end{tabular}
\caption{Dimensional model parameters. The VWF parameters, aside from L, have been estimated by fitting VWF extension in shear flow to the experimental data from \citep{Lippok2016}. 
In Sections \ref{results-behaviour} and \ref{results-artery}, the maximum VWF length is fixed at L$\,=22.6$ to match data of \citep{Schneider2007}. In Section \ref{param-sensitive} L is varied. }
\label{params}
\end{table}

\subsection{Arterial flow setup}\label{section-artery}
\begin{figure}[t!]
    \centering
\includegraphics[width=\textwidth]{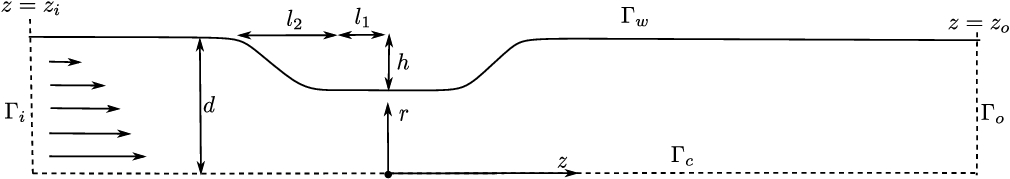}
    \caption{Diagram of $(r,z)$-plane of our axisymmetric arterial-scale stenosis geometry. Cylindrical polar coordinates $(r,z)$ are marked.  The stenosis is symmetric around $z=0$ and is defined by parameters $l_1$, $l_2$ and $h$ which define the length, steepness and height. The inlet at $z=z_i$ is $\Gamma_i$, the outlet at $z=z_o$ is $\Gamma_o$, the walls are $\Gamma_w$ and the pipe centre line is marked $\Gamma_c$.}
    \label{geom_diags}
\end{figure}

We examine our model predictions of VWF unfolding in an idealised axisymmetric stenosis under steady flow for a range of flow speeds and geometries. 
The arterial stenosis geometry is shown in Fig.~\ref{geom_diags}, the stenosis is defined by its height $h$, half-length $l_1$ and steepness $h/l_2$.  The pipe radius $d$ is chosen to match the dimensions of the coronary artery.  We solve the model in the $(r,z)$-plane assuming axisymmetry, as illustrated in Fig.~\ref{geom_diags}.
In the $(r,z)$-plane, the inlet of the pipe is located at $z=z_i$ and is denoted $\Gamma_{i}$, the outlet is located at $z=z_o$ and is denoted $\Gamma_{o}$, the pipe walls which include the stenosis are denoted $   \Gamma_{w}$ and the centre of the pipe at $r=0$ is denoted $\Gamma_{c}$. We denote the fluid flow components as $w$ and $u$ in the axial and radial directions respectively.

To close our model in this geometry we prescribe boundary conditions for \eqref{stokes} and \eqref{fene-main} as follows. The flow is driven by a unidirectional, parabolic inlet flow on $\Gamma_i$ with maximum velocity $U$. At the outlet, $\Gamma_o$, we prescribe that the normal stress is equal to a prescribed pressure, $p_a$, and that the flow is unidirectional. The latter condition creates a requirement for the pipe to be longer than any recirculation zone behind the stenosis. We prescribe no slip on the walls of the domain $\Gamma_w$.  On the centre of the domain, $\Gamma_c$, we prescribe no normal flow and a symmetry condition that the normal derivative of the axial flow vanishes.
At the inlet, $\Gamma_i$, we prescribe an inlet configuration of VWF, $\bm{A}_{\text{in}}(r)$, which is the solution of Eq.\,\eqref{fene-main} under the imposed inlet parabolic flow. Since we consider steady flow, we analyse the steady counterparts of \eqref{stokes} and \eqref{fene-main} and do not require initial conditions to close the problem.

\subsubsection{Dimensionless model}\label{section-nondim}
The dimensionless model is obtained by scaling lengths with the maximum pipe radius and the fluid velocity components with the maximum inlet velocity.  Hence we define dimensionless variables, denoted with hats, as follows
\begin{alignat}{6}
 z& = d\hat{z},&\quad r &= d\hat{r}, &\quad p &=p_a+ \frac{U\mu}{d} \hat{p},
&\quad \bm{u}&=U\hat{\bm{u}}, \label{scalings_art}
\end{alignat}
where we note that the configuration tenor is dimensionless so does not need rescaling.
The pressure scaling in \eqref{scalings_art} is defined to balance viscous forces and the pressure gradient in \eqref{stokes} relative to the prescribed outlet pressure $p_a$. The dimensionless shear rate is defined as $\dot{\gamma}=(U/d)\hat{\dot{\gamma}}$. 
Using \eqref{scalings_art} the stenosis geometry is defined by its height $\hat{h}=h/d$ and the lengths $\hat{l}_1=l_1/d$ and $\hat{l}_2=l_2/d$.
Inserting scalings  \eqref{scalings_art} into \eqref{stokes} and \eqref{fene-main} and dropping hats on dimensionless variables, we recover the dimensionless steady Navier-Stokes and FENE-P equations given by
    \begin{align}
    \na\vdot\bm{u}=0,  \quad  Re\,\bm{u}\vdot\na \bm{u}&=-\na p+\na^2 \bm{u},\label{dimless-stokes}\\
 \xi Re ( \bm{u}\vdot\na \bm{A}-\bm{A}\vdot\na \bm{u}-\na\bm{u}^T\vdot\bm{A})&=-\frac{1}{\hat{\tau}(\dot{\gamma})}(f(\bm{A})\bm{A}-a\bm{I}),\label{dimless-fene}\end{align}
  where the Reynolds number is $Re=\rho Ud/\mu$ and $\xi=\alpha\mu/d^2\rho $ is defined so that the product $\xi Re$ is the Deborah number, which represents the ratio of the timescales of protein relaxation to fluid advection. However, we choose to work with $\xi$ rather than the Deborah number so that we are able to examine the system for varying $Re.$
The FENE-P function  $f(\bm{A})=\text{L}^2/(\text{L}^2-Tr(\bm{A}))$ and $a=\text{L}^2/(\text{L}^2-Tr(\bm{I}))$ remain unchanged as L is dimensionless.
The dimensionless VWF relaxation time is \begin{align}
    \hat{\tau}(\dot{\gamma})=\frac{1}{2}\left(\tanh{\left(\hat{\beta}Re\left(\dot{\gamma}-\frac{\hat{\gamma}^*}{Re}\right)\right)}+1\right)+\delta,\label{dimless-tau}
\end{align}
where $\hat{\beta}=\beta\mu/d^2\rho$ and $\hat{\gamma}^*=\gamma^*\rho d^2/\mu$ are the dimensionless relaxation time parameters. 
The dimensionless boundary conditions for the system,  are
\begin{alignat}{2}
  w=\left(1-r^2\right), \quad u=0,  \quad \bm{A}=\bm{A}_{\text{in}}(r)\quad&\text{ on }  \quad   &\Gamma_{i}, \label{inlet_conds}\\
  \hat{\bm{n}}\vdot\bm{\sigma} \vdot\hat{\bm{n}}=0, \quad u=0 \quad&\text{ on } \quad&\Gamma_{o},\\u=0,  \quad \pdif{w}r=0
    \quad&\text{ on }   \quad  &\Gamma_{c},\label{art_vb_base}  \\   
  \bm{u}=\bm{0} \quad&\text{ on } \quad&\Gamma_{w}.\end{alignat}
In \eqref{inlet_conds} the inlet configuration of VWF, $\bm{A}_{\text{in}}(r)$, is the solution of the dimensionless Eq.\,\eqref{dimless-fene} under the imposed inlet parabolic flow.  
Dimensionless parameters and the values used in our numerical simulations in Sections \ref{results-behaviour} and \ref{results-artery} are shown in Table \ref{dimless_params}. In Section \ref{param-sensitive}, we vary the VWF unfolding parameters namely $\alpha$, $\gamma^*$, $\beta$, $\delta$, and L. For the arterial flow simulations in Section \ref{results-artery} we place the channel outlet at $\hat{z}_o=30+\hat{l}_1+\hat{l}_2$ which is sufficient to ensure the domain extends beyond the fluid recirculation zone for $\hat{h}=0.5$.
\begin{table}[t!]
\centering
\begin{tabular}{ l l l l l|}
\hline Name &
Param.    &Definition& Value(s)                        \\ 
 \hline Dimensionless pipe outlet&
$\hat{z}_o$       &    $z_o/d$   & $30+\hat{l}_1+\hat{l}_2$ \\
 Dimensionless pipe inlet& 
$\hat{z}_i$       &    $z_i/d$   & -10\\
 Dimensionless stenosis height&
$\hat{h}$       &    $h/d$   & 0.3 - 0.5
 \\
 Dimensionless stenosis length&
$\hat{l}_1$       &    $l_1/d$   & $1.5$
 \\
 Dimensionless stenosis parameter&
$\hat{l}_2$       &    $l_2/d$   & $2-5$
 
 \\ Reynolds number&
$Re$       &   $\rho U d/\mu$    & $200-500$                 \\

VWF extension parameter  &
$\hat{\beta}$   &   $\beta\mu/d^2\rho$       &    $2.16\times10^{-4}$ \\
VWF extension parameter  &
$\delta$   &   -      &    $9.7\times10^{-4}$ \\  VWF extension parameter  &
$\xi$   &   $\alpha\mu/d^2\rho$       &  0.043 \\VWF extension parameter  &
$\hat{\gamma}^*$   &   $\gamma^*d^2\rho/\mu$       &   $1.60\times10^4$
 \\
Maximum VWF length   &  L                                                                                                                               & -                   &       $22.6$                                                              \\\hline
\end{tabular}
\caption{ Dimensionless model parameters used in Sections \ref{results-behaviour} and \ref{results-artery}, those with ranges are varied, all others held fixed.}
\label{dimless_params}
\end{table}

\subsubsection{Numerical method}\label{num-low}
For an illustrative range of stenosis geometries, we consider a range of Reynolds numbers from $200$ to $500$ which produce shear rates representative of diseased arteries \citep{casa2017thrombus}. 
We solve the model using the Finite Element Method implemented using the Python Package FEniCS \citep{LoggWells2010,LoggEtal_10_2012}
which allows implementation of the weak form in the language UFL \citep{LoggEtal_11_2012,KirbyLogg2006,OlgaardWells2010}. This problem is then compiled by FIAT \citep{Kirby2004,kirby2010}. We use GMSH to construct a mesh of the stenosis geometry \citep{geuzaine2009gmsh}. Since the flow is independent of the VWF configuration, we first solve for the fluid flow and then the VWF dynamics. Full details of the numerical method are given in Appendix \ref{appendix-numerical}.

We add artificial diffusion to the FENE-P equation with a P\'{e}clet number of $10^{3}$ following the regularisation procedure commonly applied during the numerical solution of viscoelastic fluid models at high Reynolds numbers \citep{guy2014computational,sureshkumar1995effect}. Artificial diffusion allows the hyperbolic equation for 
the VWF configuration tensor to be solved using the finite element method and avoids instability at locations where the shear stress changes rapidly. The inclusion of artificial diffusion means that we must prescribe boundary conditions for the configuration tensor on all boundaries. We prescribe a symmetry condition, $\na\bm{A}\vdot\hat{\bm{r}}=\bm{0}$, on the centre of the pipe. On solid walls there are two approaches commonly used in existing numerical studies, firstly Dirichlet boundary conditions can be applied where the tensor $\bm{A}$ is set to equal the solution of \eqref{dimless-fene} in the absence of flow as in \citep{sureshkumar1995effect,paulo2014numerical}. Secondly, no normal diffusive flux can be applied on the walls as in \citep{richter2010simulations}. We adopt the latter approach as it reduces computational complexity: no diffusive flux boundary conditions can be easily applied during the Finite Element Method solution and Dirichlet conditions would require the additional solution of the FENE-P model on the walls by an alternative method. We note that since the FENE-P equation was derived in the absence of walls, the choice of boundary conditions when artificial diffusion is added is an open question for both the FENE-P model and other viscoelastic fluid models \citep{Leal1989}.

  \section{Results}
We now demonstrate how our model can be used to gain insight into VWF's behaviour in experimental flows and make predictions of the protein's dynamics in complex flow regimes. We first consider the simpler flow regimes of pure shear flow and pure elongational flow in Sections \ref{results-behaviour} and \ref{param-sensitive}, and then we consider stenotic arterial flow in Section \ref{results-artery}.

\subsection{VWF behaviour in pure shear and elongation flow}\label{results-behaviour}

In this section, we examine steady, spatially independent solutions of our VWF model in two-dimensional pure shear flow and two-dimensional pure elongational flow. 
To quantitatively compare the solutions, we set the flows to have the same scalar shear rate, $\dot{\gamma}$. We use a Cartesian coordinate system $(x,y)$ with corresponding basis vectors $(\bm{i},\bm{j})$. We note that in two-dimensional pure rotational flow with $\bm{u}=\dot{\omega}(y\bm{i}-x\bm{j})$ the solution of \eqref{fene-main} is $\bm{A}=\bm{I}$, which implies that the proteins remain at their natural length, and are randomly oriented. Hence, as expected, rotational flow only rotates the proteins but does not extend them.

We take the velocity field of the pure shear flow to be $
    \bm{u}=\dot{\gamma}y\bm{i} $, where $\dot{\gamma}$ is the shear rate. 
In two dimensions the configuration tensor has three unique components as a result of symmetry, where $A_{xx}$ and $A_{yy}$ are the average lengths squared in $x$- and $y$-directions, respectively. We seek a configuration tensor independent of time and space, which is possible since the shear rate is spatially uniform. In this case, we find that Eq.\,\eqref{fene-main} reduces to an algebraic system:
\begin{align}
 2\dot{\gamma}{\tau}A_{xy}=f(\bm{A})A_{xx}-a,\
 \dot{\gamma}{\tau}A_{yy}=f(\bm{A})A_{xy},\ f(\bm{A})A_{yy}=a.\label{shear}
\end{align}

We take the velocity field of pure elongational flow to be $
    \bm{u}=\dot{\gamma}\left(x\bm{i} -y\bm{j}\right)/{2}$, where again $\dot{\gamma}$ is the shear rate (for pure elongational flow, $\dot{\gamma}$ is sometimes referred to as the elongation rate). As in pure shear, we seek a steady, spatially independent solution of \eqref{fene-main} which gives the following algebraic system:\begin{align}
  -2\dot{\gamma}{\tau}A_{xx}=f(\bm{A})A_{xx}-a,\ 
  2\dot{\gamma}{\tau}A_{yy}=f(\bm{A})A_{yy}-a,\
    A_{xy}=0,
 \label{elong}
\end{align}
so that the configuration tensor is diagonal, reflecting that the directions of principal stretch are the  $x$- and $y$- axes.

\begin{figure}[t!]
  \centering
\includegraphics[width=.95\linewidth,valign=t]{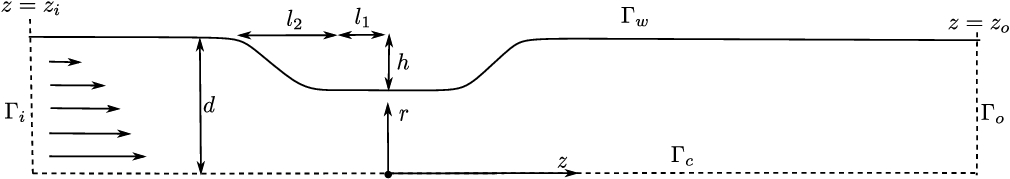}
 
  \caption{Numerical solutions of the VWF model are shown with insets $(i)$ - $(iii)$ above of typical VWF's length and alignment at three increasing shear rates. (a) In pure shear flow the protein first extends in the $x$-direction, then contracts in the $y$-direction to maintain finite length. The dashed line shows 5,096s$^{-1}$ at which VWF is half unfolded.
  (b) In pure elongation flow the protein extends in the $x$- 
direction and contracts in the $y$-direction simultaneously resulting in full unfolding at lower shear rates than in shear flow. The dashed line shows 1,947\,s$^{-1}$ at which VWF is half unfolded. VWF parameters listed in Table \ref{dimless_params} with L$\,=22.6$.}
\label{dumbell-diags}\end{figure}

The numerical solutions of \eqref{shear} and \eqref{elong} for increasing shear rate are shown in Fig.\,\ref{dumbell-diags}a and Fig.\,\ref{dumbell-diags}b respectively. For each flow type, illustrations of VWF behaviour at three increasing shear rates are shown. 
Considering first the solution in pure shear flow, Fig.\,\ref{dumbell-diags}a, we see that for values of the shear rate below the unfolding threshold, we have  $\bm{A}\approx\bm{I}$; this represents a globular protein as shown in inset $(i)$. At $\dot{\gamma}\approx2,000 $\,s$^{-1}$ the protein is only slightly unfolded, as shown in inset $(ii)$.  At large shear rates, $\dot{\gamma}\approx 5,000 $\,s$^{-1}$, the protein is 50\% unfolded and begins to align in the $x$-direction, as shown in inset $(iii)$. 
As the shear rate increases further, to maintain the finite-length restriction enforced by the VWF model, the protein's length in the $y$-direction tends to zero. We have fitted our model behaviour in shear flow to the data of \citep{Lippok2016} to obtain that at $\dot{\gamma}=5,096 $\,s$^{-1}$ the protein is unfolded to half its maximum length, which is within 1\% of the value obtained by \cite{Lippok2016} of
$5,122 $\,s$^{-1}$. The 50\% unfolding threshold is shown by the dot-dash vertical line in Fig.\,\ref{dumbell-diags}a.

In elongational flow, shown Fig.\,\ref{dumbell-diags}b, VWF remains globular for $\dot{\gamma}<100 $\,s$^{-1}$, as shown in inset $(i)$. However, at $\dot{\gamma}\approx2,000 $\,s$^{-1}$, the protein is 50\% unfolded in the $x$-direction and contracted in the $y$-direction, as illustrated in inset $(ii)$. This is in contrast to pure shear flow, where contraction in the $y$-direction only occurs at larger shear rates to maintain the protein's finite length. We predict that in elongational flow, when using L$\,=22.6$, the proteins will be 50\% unfolded at $\dot{\gamma}=1,947 $\,s$^{-1}$, which is marked on Fig.\,\ref{dumbell-diags}b by the dash-dotted line.
Since pure shear flow is the superposition of a pure elongational flow and a pure rotational flow, VWF extends to its maximum length at a much lower shear rate in pure elongational flow as rotation allows the protein to avoid unfolding. This demonstrates that our model reflects this well-established property of polymers and proteins in flow which is predicted to also occur for VWF \citep{bird1980polymer, Sing2010}.
For both pure elongational flow and pure shear flow, our modified relaxation time ensures that the proteins remain globular at low shear rates, further reflecting known VWF behaviour \cite{casa2017thrombus}.

\begin{figure}
    \centering
    \includegraphics[width=\textwidth]{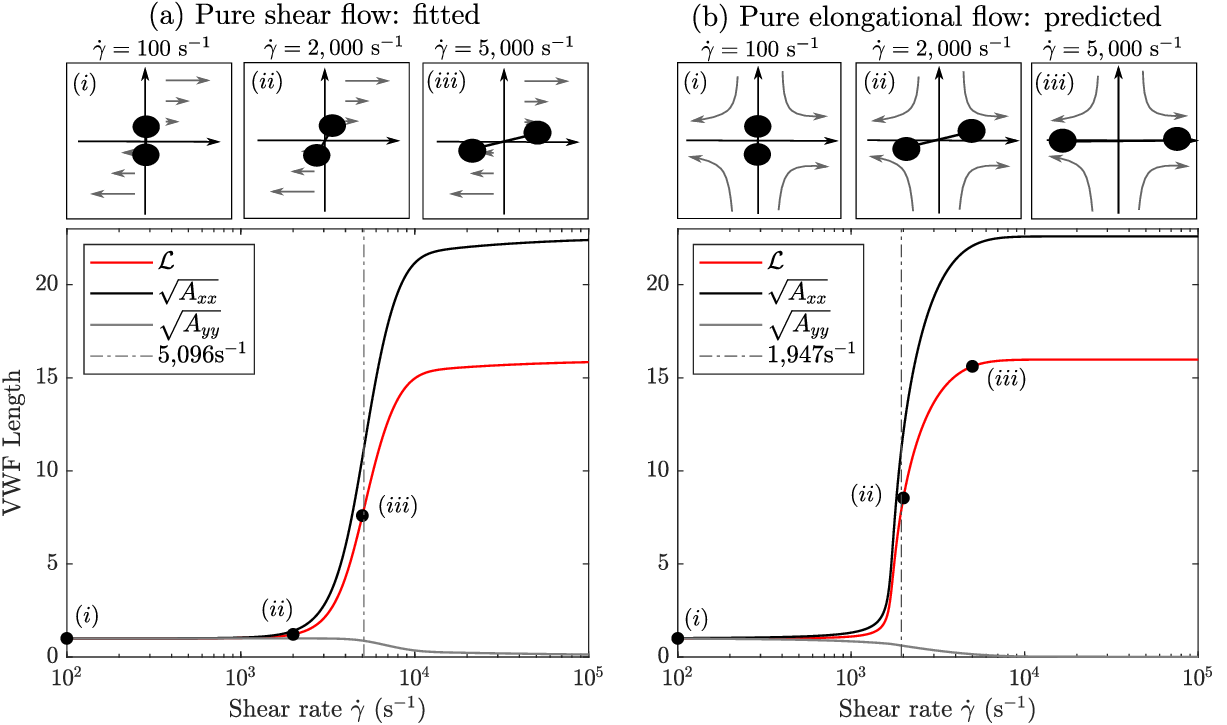}
    \caption{(a) VWF model in pure shear flow compared to \citep{Lippok2016}, the range of fitted curves obtained as L varies between 5 and 100 is shown in grey. For all values of L we are able to obtain a close quantitative match to the \cite{Lippok2016} data. (b) The corresponding range in predicted VWF behaviour in pure elongational flow is shown in grey.
 The dot-dashed line shows the minimum 50\% unfolding threshold, found with L$\,=100$, and the dashed line shows the maximum unfolding threshold which is found for L$\,=5$.
    In both plots black lines show the model solutions with L$\,=22.6$.}
    \label{fig:param_sensitive}
\end{figure}

\subsection{Pure elongational flow predictions: parameter  sensitivity}\label{param-sensitive}
In Section \ref{results-behaviour} the VWF parameters, listed in Table \ref{dimless_params}, were fitted to the data of \citep{Lippok2016} with the maximum VWF length, L, fixed at $22.6$. Using this value of L we predicted VWF will be 50\% unfolded at $\dot{\gamma}=1,947 $\,s$^{-1}$ in pure elongational flow.
Since the extent of VWF unfolding \textit{in vitro} is not well established, in this section we vary the maximum VWF length to determine the range of pure elongational unfolding rates which can be predicted by our model. 

For L between 5 to 100, the best fit of the model to data is calculated using the data of \citep{Lippok2016}. The shaded region in Fig.\,\ref{fig:param_sensitive}a shows the range in fitted behaviour as L varies, the fitting used in Sections \ref{results-behaviour} and \ref{results-artery} is shown by the black line. For all L values, we are able to obtain a mean error within 2\% of the \citep{Lippok2016} data in pure shear flow. 

The predicted behaviour in pure elongational flow is shown in Fig.\,\ref{fig:param_sensitive}b. The shaded region represents the solution evaluated using the best fit of parameters from Fig.\,\ref{fig:param_sensitive}a.
The predicted 50\% unfolding threshold in pure elongational flow varies between approximately $\dot{\gamma}=635 $\,s$^{-1}$ to $3,280 $\,s$^{-1}$. The smallest unfolding threshold of $635 $\,s$^{-1}$ is obtained when the largest value of L$\,=100$ is used, showing that the proteins which are capable of sustaining very large extensions also unfold at lower shear rates. The significant variability in the pure elongational flow thresholds demonstrates a large degree of sensitivity in the model output to the value of L selected and further motivates the need to experimentally quantify the extension VWF is able to sustain in flow.

\subsection{VWF behaviour in \textit{in vivo} flow}
\label{results-artery}  
\begin{figure}[!h]
    \centering

\includegraphics[width=0.99\textwidth]{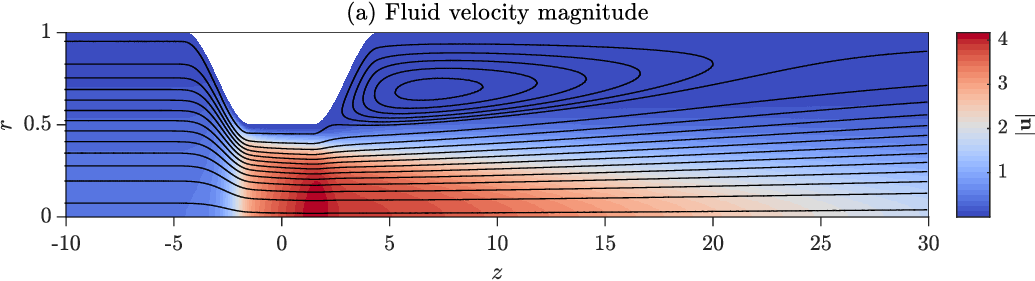}

\includegraphics[width=0.99\textwidth]{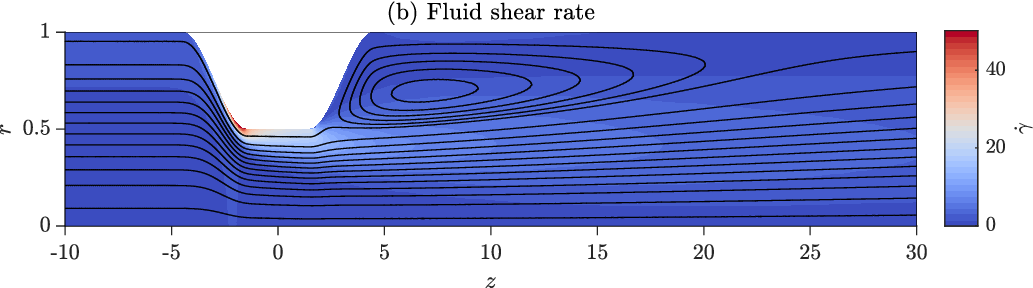}

\includegraphics[width=0.99\textwidth]{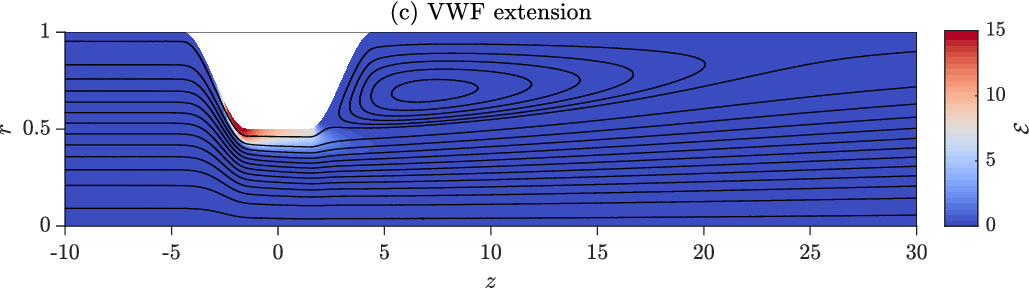}
    
    \caption{Dimensionless numerical solutions for $Re=400$ overlayed by fluid streamlines.\newline (a) Fluid velocity magnitude. A recirculation zone forms downstream of the stenosis, indicated by circular streamlines. (b) Fluid shear rate, which is greatest at the leading edge of the stenosis. \newline (c) VWF extension. The proteins are most extended by the stenosis wall and are fully extended at the leading edge of the stenosis. Parameter values: $\hat{l}_1=1.5,\ \hat{l}_2=2,\ \hat{h}=0.5$, L$\,=22.6.$} \label{250}
\end{figure}

We now examine the model's predictions for VWF's behaviour in steady stenoic arterial flow.
Figure \ref{250} shows the dimensionless numerical solution of the model obtained for $Re=400$. All subfigures illustrate solutions overlaid by the fluid closed streamlines. 
At this Reynolds number, a recirculation zone forms downstream of the stenosis as illustrated by the streamlines. The magnitude of the fluid velocity is shown in Fig.\,\ref{250}a. The flow is four times faster as it crosses the stenosis compared to upstream. The fluid shear rate, shown in Fig.\,\ref{250}b, is greatest at the leading edge of the stenosis at $z=-2$ where it reaches $\dot{\gamma}\approx 55$. The shear rate is much lower away from the boundary and in the flow recirculation zone. 
VWF extension $\mathcal{E}$ is shown in Fig.\,\ref{250}c.  VWF reaches $\mathcal{E}\approx15$ which is the maximum extension achievable with a maximum VWF length of L$\,=22.6$. The maximum extension is obtained at the leading edge of the stenosis at $z=-2$ where $\dot{\gamma}$ is the greatest.

   \subsubsection{The effect of Reynolds number on VWF unfolding}\label{results-re}
We now examine how VWF extension changes as the Reynolds number varies for a fixed stenosis geometry. In this section to compare the shear rate obtained at the boundary for different flow rates we define the scaled wall shear rate (WSR) as the dimensionless shear rate multiplied by the Reynolds number, $Re \dot{\gamma}$, this remains dimensionless but reflects how the magnitude of the dimensional shear rate changes as the flow rate increases. 

\begin{figure}[!t]    \centering
 \includegraphics[width=\textwidth]{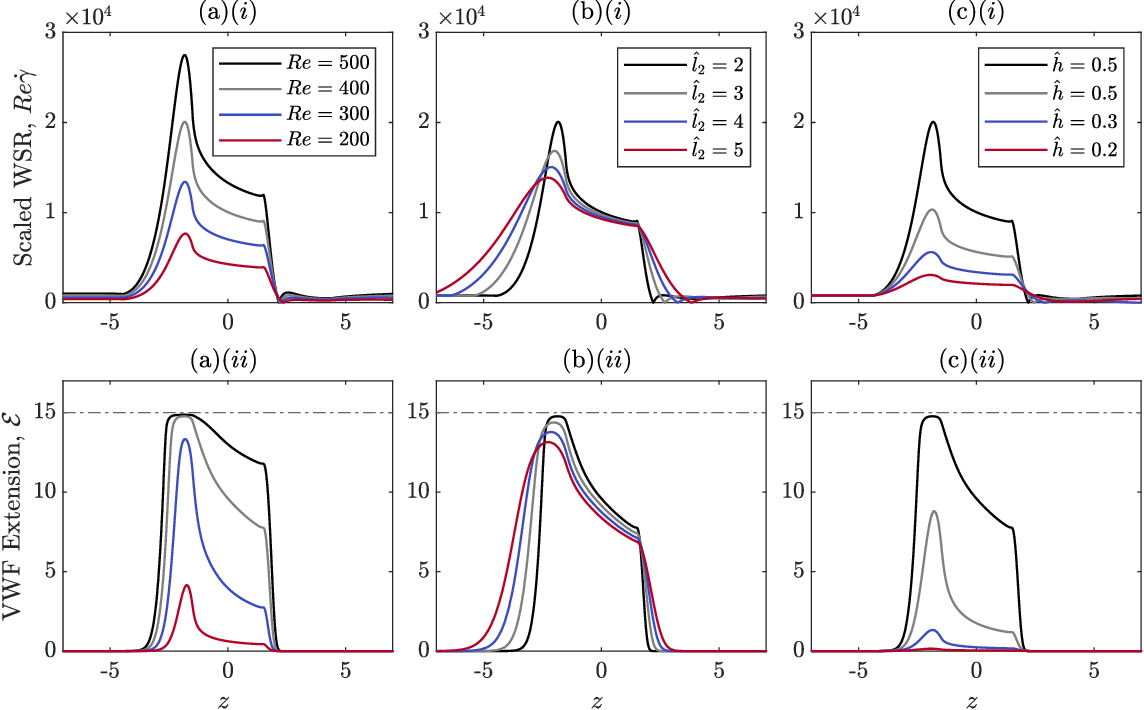}

\caption{ VWF extension for varying stenosis geometry and Reynolds numbers. When not stated in the legend, all other geometry parameters are: $\hat{l}_1=1.5$, $\hat{l}_2=2$, $\hat{h}=0.5$, L$\,´=22.6$ and $Re=400$. Columns show behaviour for (a) 
 increasing $Re$, (b) increasing stenosis steepness and (c) increasing stenosis height respectively. Subfigures $(i)$ show the scaled wall shear rate (WSR), defined as $Re\dot{\gamma}$, and the corresponding subfigures
$(ii)$ show the VWF extension obtained at the wall.
  }\label{parameter_test}\end{figure}

The scaled wall shear rate on the stenosis wall for $Re$ from ranging 200 to 500 is shown in Fig.\,\ref{parameter_test}a$i$, illustrating that as the Reynolds number increases, the shear rate increases. The maximum shear rate occurs at the leading edge of the stenosis for all $Re$.
VWF extension on the stenosis wall is shown in Fig.\,\ref{parameter_test}a$ii$.
For all Reynolds numbers, the maximum extension is obtained at the point on the stenosis wall where the wall shear rate is greatest. As $Re$ increases VWF extends more at the stenosis wall as a result of the increasing shear rate. Furthermore, the nonlinear dependence of VWF extension on the shear rate is demonstrated as the protein reaches an extension of nearly 100\% at $Re=400$ but only 33\% at $Re=200$. 

   \subsubsection{Stenosis geometry and VWF extension}\label{results-geom}

  In Section \ref{results-re} we found that the greatest VWF extension is obtained at the wall, hence we now examine how varying stenosis geometry alters the value and axial position of this extension for fixed Reynolds number of $Re=400$.
The scaled wall shear rate and VWF extension for increasing stenosis steepness are shown in Fig.\,\ref{parameter_test}b$i$ and Fig.\,\ref{parameter_test}b$ii$. We increase the stenosis steepness by decreasing the parameter $\hat{l}_2$. 
Increasing the steepness of the stenosis slightly increases the maximum shear rate in the pipe, causing the VWF to unfold more. However, for steeper stenoses the increased shear rates, and correspondingly VWF extension, are confined to a smaller region. 

The wall shear rate and VWF extension for increasing stenosis height are shown in Fig.\,\ref{parameter_test}c$i$ and Fig.\,\ref{parameter_test}c$ii$.
 Increasing the stenosis height drastically increases the maximum shear rate in the pipe and causes VWF to unfold to a greater extent. For smaller stenoses with $h\leq0.2$ a fluid recirculation zone does not form since the shear rate $\dot{\gamma}>0$ for all $z>0$. The absence of a recirculation zone means that there will be more significant transport of VWF behind the stenosis which could alter thrombus location.

\subsubsection{Elongational flow structures in arteries and VWF unfolding}\label{results-flow}
\begin{figure}[!t]
    \centering    
   \includegraphics[width=0.95\textwidth]{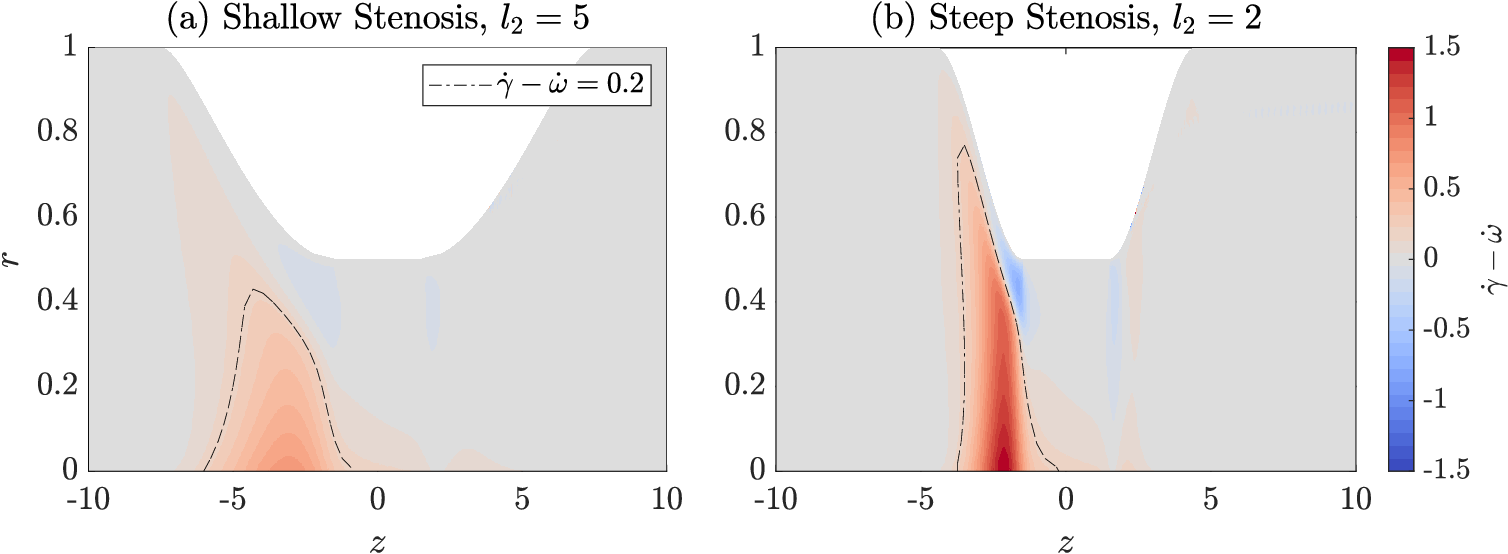}
    \caption{ Flow structures over the stenosis. The flow is elongational when $\dot\gamma-\dot\omega\ll0$ and rotational when $\dot\gamma-\dot\omega\ll0$. Flow over: (a) a shallow stenosis with $\hat{l}_2=5$ and (b) a steeper stenosis with $\hat{l}_2=2$. For the two cases the region where $\dot\gamma-\dot\omega=0.2$  is shown by the dashed black line, the interior defines regions in which we have significantly more elongation than rotation. The elongational flow (b) has a much larger disparity between the shear rate and the rotation rate at the entrance to the stenosis. Both have: $\hat{l}_1=1.5,\ \hat{h}=0.5,\ Re=400$.} \label{class}
\end{figure}

In Section \ref{results-geom}, we showed that increasing the steepness of the stenosis alters the flow, leading to a higher wall shear rate. Fig.\,\ref{class} shows the difference between the shear rate and the rotation rate, $\dot\gamma-\dot{\omega}$, for a steep stenosis compared to a more shallow stenosis, with red regions on Fig.\,\ref{class} showing regions of elongational flow and blue regions showing rotational flow. The steeper stenosis leads to elongational flows with $\dot\gamma-\dot{\omega}$ three times larger than the shallow stenosis. 

To highlight this we show regions for which $\dot\gamma-\dot{\omega}=0.2$, by the dashed regions in Fig.\,\ref{class}. The interior of this line defines regions where the flow is highly elongational.
The maximum shear rate obtained in these highly elongational regions is $\dot\gamma=1.3$ and  $\dot\gamma=3.7$  for the shallow stenosis and steep stenosis, respectively. These correspond to $\dot\gamma=317.1\,\text{s}^{-1}$ and  $\dot\gamma=902.0\,\text{s}^{-1}$ in dimensional terms. 
For L$\,=22.6$ our model predicted that in pure elongational flow VWF is half unfolded at $1,947\,\text{s}^{-1}$ whereas VWF is half unfolded at $5,096\,\text{s}^{-1}$ in pure shear flow. Since the flow in the centre region of the stenotic artery is not pure elongational flow, we expect that the unfolding threshold in this region will be larger than $1,947\,\text{s}^{-1}$ but still smaller than the pure shear flow unfolding threshold. In the highly elongational regions, the shear rate does not reach the pure elongational flow threshold of $1,947\,\text{s}^{-1}$. As a result, VWF only unfolds to 2.3\% and 0.7\% of its maximum length in the indicated regions in Fig.\,\ref{fig:flow_structure}. This is in contrast to the extension achieved at the wall where VWF can reach extensions of 98\% and 88\% in the steep and shallow stenosis cases respectively. The lack of significant unfolding in the elongational flow region is in contrast to the work of \cite{zhussupbekov2021continuum} where the authors found that VWF will be fully extended in the centre of the flow. 

The predicted significance of elongational flow on VWF unfolding depends on model parameterisation and the predicted unfolding rate in pure elongational flow. 
In Section \ref{param-sensitive}, we demonstrated that our predicted unfolding thresholds in pure elongational flow ranges from approximately $600-3,200\,\text{s}^{-1}$ depending on the maximum VWF length L, which is not known. The smallest unfolding threshold was found for proteins with the largest maximum lengths L$\,=100$.  This suggests that if we have a small unfolding threshold then we could see proteins reach up to 50\% extension away from the wall in elongational flow regions.

\section{Discussion}\label{section-discussion}
In this paper we have presented a model for the dynamics of shear-sensitive blood protein VWF using a dilute limit of the viscoelastic fluid model FENE-P with a modified relaxation time. The modified relaxation time captures VWF propensity to unfold in response to the fluid shear rate. This is characterised using parameter estimates from the experimental data of \citep{Lippok2016}. Through a configuration tensor, our model can describe VWF’s length and orientation in any combination of elongational, shear and rotational flows, defined as $\dot{\gamma}\gg\dot{\omega}$, $\dot{\gamma}\approx\dot{\omega}$ and $\dot{\gamma}\ll\dot{\omega}$ respectively.
 Using an idealised arterial stenosis geometry, we demonstrated that increasing the fluid flow rate and stenosis height have the strongest effect on the wall shear rate and therefore VWF’s extension at the wall. Since platelets are transported in large quantities in the cell-free layer by the wall, VWF molecules which are extended close to the arterial wall will most readily bind with platelets to form a thrombus \cite{casa2017thrombus}. 

Our model is able to reproduce the dependency of VWF behaviour on the flow structure, namely that the protein unfolds at lower shear rate in pure elongational compared to pure shear flow \newline\citep{babcock2003visualization,smith1999single}.  Our prediction of the shear rate at which VWF unfolds in pure elongational flow varies depending on the value of maximum extension which VWF can achieve, which is not definitively established in the literature.  In our model the parameter L restricts the value of VWF extension. For L$\,=5$ - $100$, VWF can extend to at most $\approx 2-70$ times its natural length. For this range of the maximum VWF length we estimate that VWF will be 50\% unfolded between $600$ - $3,200\,\text{s}^{-1}$ in pure elongational flow. This agrees with existing discrete models of VWF which uniformly estimate that VWF unfolds at a lower shear rate in elongational flow compared to shear flow. Furthermore, our estimated range of the unfolding threshold in pure elongational flow falls within the range of values predicted by discrete mathematical models of single VWF molecules: 500$\,\text{s}^{-1}$ \citep{Sing2010}, 2,400$\,\text{s}^{-1}$ \citep{nguyen2021unraveling}, 2,500$\,\text{s}^{-1}$ \citep{kania2021predicting} and 3,500 $\,\text{s}^{-1}$ \citep{dong2019mechano}. 

This model is able to examine VWF behaviour in the complex, multidimensional flows which occur in diseased arteries. We show VWF is most unfolded in the shear flow close to the stenosis wall, with the maximum extension occurring at the leading edge of the stenosis. This provides patterns of elongation of the protein along the wall which could be combined with a model of platelet transport to predict thrombus formation. We have shown that elongational flow occurs within stenosed geometries, with the difference between the shear rate and the rotation rate increasing as the steepness of the stenosis increases. Our model can evaluate the degree to which VWF unfolds in free flow away from the wall compared to the wall extension. Using a single value of VWF extension which matches the data of \cite{Schneider2007}, namely L$\,=22.6$, our model predicts VWF only reaches 2\% of its maximum length in the highly elongational flows away from the wall where the maximum shear rate is 171$\,\text{s}^{-1}$. However, our parameter sensitivity analysis suggests that there are some parameter regimes, depending on the value of L selected, in which significant unfolding could be found away from the wall.  

The structure of our model differs from the only continuum model of VWF to date by \cite{zhussupbekov2021continuum}. \cite{zhussupbekov2021continuum} uses experimental data from DNA unfolding to define regions where the flow is sufficiently elongational to unfold VWF \citep{babcock2003visualization}. \cite{zhussupbekov2021continuum} then enforce that the proteins unfold at 500$\,\text{s}^{-1}$ in these regions of elongational flow. Using these parameter choices \citep{zhussupbekov2021continuum} predict that VWF will fully unfold in the flow away from the wall in microfluidic stenosis simulations. Our model does not include a threshold at which the flow is classified as elongation; instead, the \textit{flow structure is encoded }in \eqref{fene-main} through the deformation tensor. The deformation tensor is then combined with a single constitutive relaxation time which models VWF's ability to unfold. This allows our model to be easily parameterised using data from shear flow, eliminating the need to rely on data obtained for other proteins, which may not be accurate for VWF. 

The accuracy of our predictions relies on the estimation of the model parameters which describe VWF’s unfolding through the nonlinear relaxation time $\tau$. We estimated these parameters, aside from VWF length $L$, by comparing our model predictions in shear flow to the data of \citep{Lippok2016}. This required the estimation of five unknown parameters. Our estimate yields a 1.82\% error in the relative length of VWF compared to \citep{Lippok2016}.  However, this estimation was done using a single minimsation algorithm, and it is possible that alternative minima could exist which yield a better fit to the Lippok data. Finally, in this paper we varied the maximum VWF length to determine the variation in best fit obtained to the \cite{Lippok2016} data. The resulting predicted behaviour in pure elongational flow varied significantly over the range of L$\,=5-100$. When further data is available for the maximum extension of VWF in free flow the model parameters which determine VWF unfolding can be readily updated allowing the model to more precisely estimate the elongational flow behaviour of VWF.

There are several limitations and possible extensions of the theoretical framework of our model which we now detail. Firstly, our model does not include any history effects, for instance, the proteins do not require exposure to high shear stresses for a certain period of time to unfold. Furthermore, our model does not include the hysteresis of VWF, whereby the proteins relax back to their original length over a longer timescale than extension. This would mean that the proteins could remain unfolded downstream of the stenosis which could be significant for thrombus formation behind the stenosis. Our model could be extended to include hysteresis by following the construction in \citep{zhussupbekov2021continuum} and categorising the proteins as extending, which unfold rapidly, and retracting, which refold more slowly. However, this would require formulating how proteins move between the two categories, adding significant complexity to the model in physiological flows. 

There are several theoretical extensions to our modelling framework which would improve its ability to describe VWF when in close proximity to the artery wall. In this paper we used the solution of the FENE-P equation to describe VWF length when at the vessel wall. However, the FENE-P equation is derived for a protein in the absence of walls.  The effect of walls has been included in similar non-Newtonian models of confined flows of proteins \citep{biller1987flow} and confined flow of bacteria \citep{saintillan2015theory}. 
  However, this introduces reflection conditions or binding conditions on the probability density function from which the configuration tensor is derived. This adds complexity to the model construction as the arising equation for the configuration tensor does not have a closed form \citep{biller1987flow}.  
  VWF unfolding behaviour when tethered to a non-reactive wall differs significantly from its behaviour in free flow, so it is not clear if the unfolding relation fitted in shear flow used in this paper would effectively describe the dynamics of VWF when close to or bound to a wall \citep{Fu2017}. Finally, when binding to a reactive wall VWF has been shown to form bundles or carpets of tangled proteins \citep{Schneider2007,colace2013direct}; since the FENE-P equation describes dilute suspensions of polymers or proteins our model would not be able to capture the dynamics of dense suspensions. Insights from discrete models of VWF could be used to effectively determine how best to include the effects of binding or protein-protein interactions into a continuum framework \citep{liu2022sipa,wang2019shear}.

In this paper we examine flow and VWF dynamics within arterial scale stenoses, as this is the most clinically relevant scale and geometry at which high shear thrombosis occurs. 
However, VWF-mediated thrombosis can also occur at the location of an arterial stent or on a prosthetic heart valve \citep{casa2017thrombus}. Our model can be readily applied to examine these alternative geometries or indeed any vessels or devices in which the continuum approximation for the VWF suspended in blood is valid. This holds when the vessel diameter is significantly larger than the radius of a red blood cell (approximately 3.5$\,\mu$m \citep{colace2013direct}). As a result, our model can be applied in smaller vessels such as arterioles or in microfluidic devices which are regularly used to study thrombosis \textit{in vitro} \citep{westein2013atherosclerotic,liu2022sipa}. 

\section{Conclusion}\label{section-conclusion}

In this paper we have presented a novel continuum model to describe the dynamics of VWF in blood.  Our model uses a single constitutive relation to describe VWF’s propensity to unfold at a given shear rate which is parameterised to match experimentally measured VWF behaviour in shear flow. The model is then able to quantitatively predict VWF length and orientation in any combination of flow types which occur in diseased arteries. Crucially, our model can examine VWF dynamics in elongational flows which are challenging to examine experimentally and which are predicted to facilitate excessive VWF unfolding. Our model could be readily incorporated into a continuum model of high-shear thrombosis where the configuration tensor can be used to mechanistically describe VWF's transport and binding as a function of its conformation.

\section{Data accessibility}
Files for numerical solution and figure production can be found in the repository \newline{https://github.com/Edwina-Yeo/VWF-Modelling}.
\section{Statements and Declarations}
We declare we have no competing interests.
This work was supported by an EPSRC Studentship and an EPSRC Doctoral Prize award (project reference: 2100104) (E.Y.).

\bibliography{main}        
\bibliographystyle{apacite}

\section{Appendix}
\subsection{Parameter estimation}\label{appendix-param}
We estimate the model parameters which determine VWF's unfolding namely $\alpha$, $\beta$, $\delta$ and $\dot{\gamma}^*$, using the empirical VWF cleavage rate of \cite{Lippok2016} normalised using the maximum cleavage rate obtained in their work ($3.5 \times 10^{-3}$nM/s). Following the assumption of \cite{Lippok2016} that the proteins cleave at a rate proportional to their length, this normalised cleavage rate represents the normalised VWF extension. We use the solution of the FENE-P equation in two-dimensional shear flow at $\dot{\gamma}_i$, which we denote as $\bm{A}(\dot{\gamma}_i)$ to calculate the extension at that shear rate $\mathcal{E}(\dot{\gamma}_i)$ using \eqref{L_def}. We use this extension to define the mean error made to \cite{Lippok2016} data as
\begin{align}
  E=\frac{1}{N}\sum_{i=1}^{N}|\tilde{\mathcal{E}}(\dot{\gamma}_i)- \mathcal{E}(\dot{\gamma}_i)|,\label{E_def}
\end{align}
where in practice we use $N=400$ discrete values of the shear rate between $\dot{\gamma}=1\,\text{s}^{-1}$ and $\dot{\gamma}=10^5\,\text{s}^{-1}$ equally spaced on a log scale.
We use the gradient-based minimiser \textit{fmincon} from MATLAB's optimisation toolbox to determine the VWF parameters which minimise $E$. We use the initial guess of $\beta=0.01\,\text{s}$ $\dot{\gamma}=10^4\,\text{s}^{-1}$, $\alpha=0.01\,\text{s} $ and $\delta=10^{-4}$.

We first seek a set of optimal parameters with the maximum VWF length fixed at L$\,=22.6$ which is set so that the maximum extension VWF can achieve in two-dimensions matches the maximum extension measured by \citep{Schneider2007}.
We set bounds on the minimisation so that we seek optimal parameters which satisfy $10^{-8}<\delta<10^{-3}$, $10^{-8}\,\text{s}<\beta<10^{-2}\,\text{s}$, $10^{-6}\,\text{s}<\alpha<0.1\,\text{s}$, $4000\,\text{s}^{-1}<\dot{\gamma}^*<2\times10^4\,\text{s}^{-1}$, which are motivated by the physical role of each parameter in the relaxation time. We find that the parameter values listed in Table \ref{params} minimise the error obtained compared to \citep{Lippok2016} with $E=    0.0182
$ as defined in \eqref{E_def}, which is equivalent to an average percentage error of 1.82\%. 

In Section \ref{param-sensitive} we vary the value of L used. We use seven values of L between 5 and 100, obtaining a best estimate of the parameters: $\alpha$, $\beta$, $\delta$ and $\dot{\gamma}^*$, in each case using numerical continuation. We keep the bounds on the parameter space unchanged in this process.

\subsection{Numerical scheme and validation}\label{appendix-numerical}
We use the package FEniCS version \textit{2019.2.0.dev0} and code construction is based on examples in  \citep{AlnaesBlechta2015a,LoggEtal2012} for the solution of Stokes equations and advection-diffusion equations.

In our numerical solution of the steady Navier-Stokes equations \eqref{dimless-stokes} we use Taylor Hood elements of first- and second-order for the pressure and velocity vector, respectively. The steady nonlinear system is solved using the inbuilt Newton Solver \textit{solve} as part of the FEniCS package. The velocity gradients in each direction, along with the wall shear rate on the pipe wall, are determined using first-order elements as functions of the velocity solution. The velocity gradients and velocity field are then used to solve the modified FENE-P equation \eqref{dimless-fene}.

 The FENE-P equation consists of four coupled advection-diffusion equations with the velocity field given by the solution of the Navier-Stokes equations. As discussed above, we include diffusion in these equations for numerical tractability. We use first-order Lagrange elements to solve for each component of the configuration tensor. To solve the system for each component of the configuration tensor, $\bm{A}$, we employ continuation in the Reynolds number, with unit steps performing well. Our initial guess for the solver for $Re=0$ is that $\bm{A}=\bm{I}$.  The inlet value of the configuration tensor is found by solving Eq.\,\eqref{dimless-fene} under the imposed inlet flow for which we solve the nonlinear system numerically, using the \textit{NumPy} Newton solver \textit{fsolve} without numerical continuation, with an initial guess of $\bm{A}=\bm{I}$.
\begin{figure}[t]    \centering
 
  \centering
  \includegraphics[width=0.95\textwidth]{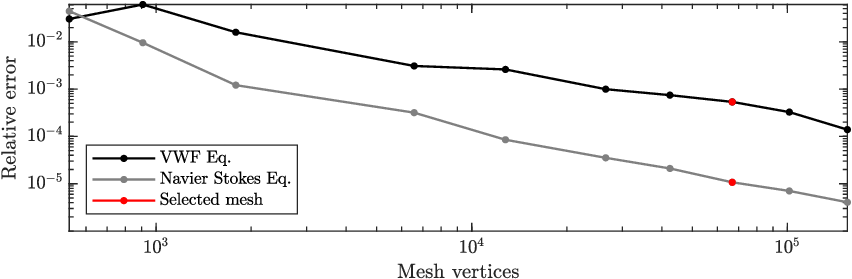}

\caption{ (a) Relative error in maximum extension of the FENE-P equation and maximum fluid velocity for increasing numbers of mesh vertices. (b) Convergence of the minimum value of VWF extension to zero. Selected mesh shown in red. Parameters:  $\hat{h}=0.5, \hat{l}_1=1.5,\ \hat{l}_2=2$, $Re=500$, L$\,=22.6$.  }\label{convergence}\end{figure}

We construct a mesh in GMSH which is finer closer to the boundary and at the upstream edge of the stenosis where the shear rate is greatest. In Fig.~\ref{convergence}a we compare the maximum value of VWF extension and fluid velocity magnitude obtained on a sequence of meshes with increasing numbers of mesh vertices to the same quantities obtained using a fine reference mesh with $ 6\times10^5$ vertices. We use the mesh that has approximately $ 6.7\times10^4$ vertices on which we achieve a maximum relative error of 0.1\% on all variables compared to the aforementioned reference mesh. 
We define the relative error by dividing the difference between the value on the fine reference mesh and the coarse mesh by the maximum absolute value on the fine mesh. 

The steep gradients in the fluid shear rate can lead to spurious oscillations in the components of the configuration tensor $\bm{A}$. To ensure that our mesh is suitably fine to capture the large gradients in the tensor components without oscillations we examine the minimum value of the trace of the configuration tensor obtained on each mesh. The trace $\bm{A}$ should be bounded below by one. Our selected mesh gives a relative error, defined as  $(Tr(\bm{A})-1)/(L^2-1)$, equal to -0.0022 hence the amplitude of any oscillations in the configuration tensor components is less than 1\% when compared to the maximum extension. 

\newpage

\end{document}